\documentclass[10pt,a4paper]{article}

\usepackage[T1]{fontenc }
\usepackage{amsmath}
\usepackage{amscd}
\usepackage{amsfonts}
\usepackage{amssymb}
\usepackage{bm}
\usepackage{graphicx}
\usepackage{geometry}

\def\1{{\mathchoice{\rm 1\mskip-4mu l}{\rm 1\mskip-4mu l}%
{\rm 1\mskip-4.5mu l}{\rm 1\mskip-5mu l}}}

\newcommand{\ot}{{\,\otimes\,}}
\newcommand{{\Cd}}{{\mathbb{C}^d}}

\def\oper{{\mathchoice{\rm 1\mskip-4mu l}{\rm 1\mskip-4mu l}%
{\rm 1\mskip-4.5mu l}{\rm 1\mskip-5mu l}}}
\def\<{\langle}
\def\>{\rangle}

\title{Entanglement witnesses for $d \otimes d$ systems
and new classes of entangled qudit statess}
\author{ D. Chru\'sci\'nski and A. Rutkowski\\[1ex]
Institute of Physics, Nicolaus Copernicus University \\
Grudziadzka 5, 87--100 Toru\'n, Poland}
\date{}
\begin{document}
\maketitle

\abstract{ We provide a new class of entanglement witnesses for $d
\ot d$ systems (two qudits). Our construction generalizes the one
proposed recently by Jafarizadeh et al. for $d=3$ and $d=4$ on the
basis of semidefinite linear programming. Moreover, we provide a new
class of PPT entangled states detected by our witnesses which
generalizes well known family of states constructed by Horodecki et
al. for $d=3$.}

\section{Introduction}

In recent years, due to the rapid development of quantum information
theory \cite{QIT} the necessity of classifying entangled states as a
physical resource is of primary importance. It is well known that it
is extremely hard to check whether a given density matrix describing
a quantum state of the composite system is separable or entangled.
There are several operational criteria which enable one to detect
quantum entanglement (see e.g. \cite{Horodecki-review} for the
recent review). The most famous Peres-Horodecki criterion  is based
on the partial transposition: if a state $\rho$ is separable then
its partial transposition $\rho^\Gamma = (\oper \ot {\rm T})\rho$ is
positive. States which are positive under partial transposition are
called PPT states. Clearly each separable state is necessarily PPT
but the converse is not true. We stress that it is easy to test
wether a given state is PPT, however, there is no general methods to
construct PPT states.

The most general approach to characterize quantum entanglement uses
a notion of an entanglement witness (EW) \cite{HHH,Terhal1,Terhal2}.
A Hermitian operator $W$ defined on a tensor product
$\mathcal{H}=\mathcal{H}_1 \ot \mathcal{H}_2$ is called  an EW iff
1) $\mbox{Tr}(W\sigma_{\rm sep})\geq 0$ for all separable states
$\sigma_{\rm sep}$, and 2) there exists an entangled state $\rho$
such that $\mbox{Tr}(W\rho)<0$ (one says that $\rho$ is detected by
$W$). It turns out that a state is entangled if and only if it is
detected by some EW \cite{HHH}. There was a considerable effort in
constructing and analyzing the structure of EWs
\cite{O}--\cite{iran}. However, there is no general method to
construct such objects.

The simplest way to construct EW is to define $W = P + Q^\Gamma$,
where $P$ and $Q$ are positive operators. It is easy to see that
$\mbox{Tr}(W\sigma_{\rm sep})\geq 0$ for all separable states
$\sigma_{\rm sep}$, and hence if $W$ is non-positive, then it is EW.
Such EWs are said to be decomposable \cite{O}. Note, however, that
decomposable EW cannot detect PPT entangled state  and, therefore,
such EWs are useless in the search for bound entangled state.
Unfortunately,  there is no general method to construct
non-decomposable EW (nd-EW) and only very few examples of nd-EWs are
available in the literature.

In the present paper we provide a class of nd-EWs for $d \ot d$
systems. Our construction  generalizes the one proposed recently by
Jafarizadeh et al. \cite{iran} for $d=3$ and $d=4$ on the basis of
semidefinite linear programming. Moreover, we provide a new class of
PPT entangled states detected by our witnesses which generalizes
well known family of states constructed by Horodecki et al. for
$d=3$ \cite{HOR}.

The paper is organized as follows: in the next section we introduce
a class of circulant operators \cite{CIRCULANT} (see also \cite{Art}
for more abstract discussion). It tirns out that circulant operators
defines a natural arena for constructing interesting classes of
bi-partite quantum states and the corresponding EWs. Section
\ref{EWS} provides the basic construction of EWs. Then in section
\ref{PPT} we show that our witnesses are non-decomposable by
providing a family of PPT entangled states detected by our
witnesses. A brief discussion is included in the last section.

\section{Circulant operators for two qudits}  \label{CIRC}

Consider a class of linear Hermitian operators in $\mathbb{C}^d \ot
\mathbb{C}^d$ constructed as follows: let $\{|0\>,\ldots,|d-1\>\}$
denotes an orthonormal basis in $\mathbb{C}^d$ and let $S :
\mathbb{C}^d \rightarrow \mathbb{C}^d$ be a shift operator defined
as follows
\begin{equation}\label{}
    S|k\> = |k+1\> \ ,\ \ \ ({\rm mod}\ d) \ .
\end{equation}
One introduces the following family of $d$-dimensional subspaces in
$\mathbb{C}^d \ot \mathbb{C}^d$:
\begin{equation}\label{}
    \Sigma_0 = {\rm span}\{ |00\>, \ldots, |d-1,d-1\>
    \} \ ,
\end{equation}
and
\begin{equation}\label{}
    \Sigma_n = (\mathbb{I} \ot S^n) \Sigma_0 \ , \ \ n=1,\ldots,d-1\
    .
\end{equation}
It is clear that $\Sigma_m$ and $\Sigma_n$ are mutually orthogonal
for $m \neq n$ and hence the collection $\{ \Sigma_0, \ldots ,
\Sigma_{d-1}\}$ defines direct sum decomposition of $\mathbb{C}^d
\ot \mathbb{C}^d$
\begin{equation}\label{dd-Sigma}
\mathbb{C}^d \ot \mathbb{C}^d  = \Sigma_0 \oplus \ldots \oplus
\Sigma_{d-1}\ .
\end{equation}
Following \cite{CIRCULANT} we call (\ref{dd-Sigma}) a {\em circulant
decomposition}.  Now we construct a circulant Hermitian operator
corresponding to the circulant decomposition (\ref{dd-Sigma}). Let
us introduce a set of Hermitian   $d \times d$ matrices $a^{(n)} = [
a^{(n)}_{ij}]\, ; \, n=0,1,\ldots,d-1$, and define Hermitian
operators $A_n$ supported on ${\Sigma}_n$ via the following formula
\begin{eqnarray}\label{}
A_n &=&  \sum_{i,j=0}^{d-1}\, a^{(n)}_{ij}\, |i\>\<j| \ot S^n\, \,
|i\>\<j|\, S^{\dagger \, n} \nonumber \\
&=&  \sum_{i,j=0}^{d-1}\, a^{(n)}_{ij}\, |i\>\<j| \ot |i+n\>\<j+n|\
.
\end{eqnarray}
Finally, we define the circulant Hermitian operator
\begin{equation}\label{}
    A = A_0 + A_1 + \ldots +
    A_{d-1}\ .
\end{equation}
Note, that if all $A_n$ are semipositive definite and ${\rm Tr}\, A
=1$, then $A$ defines a legitimate quantum state of two qudits
called {\em circulant state} \cite{CIRCULANT}. Interestingly, many
well known examples of quantum states of composite $d \ot d$ systems
belong to this class (see \cite{CIRCULANT,Art} for examples).

The crucial property of circulant operators is based on the
following observation: the partially transposed circulant operator
$A$ displays similar circulant structure, that is,
\begin{equation}\label{}
    A^\Gamma  = \widetilde{A}_0 \oplus \ldots
    \oplus \widetilde{A}_{d-1} \ ,
\end{equation}
where the Hermitian operators $\widetilde{A}_n$ are supported on the
new collection of subspaces $\widetilde{\Sigma}_n$ which are defined
as follows:
\begin{equation}\label{}
    \widetilde{\Sigma}_0 = {\rm span}\{ |0,\pi(0)\>, |1,\pi(1)\>, \ldots,
    |{d-1},\pi(d-1)\>     \} \ ,
\end{equation}
where $\pi$ is a permutation defined by
\begin{equation}\label{pi}
    \pi(k) = d-k \ , \ \ \ ({\rm mod} \ d)\ .
\end{equation}
The remaining subspaces $\widetilde{\Sigma}_n$ are defined by a
cyclic shift
\begin{equation}\label{}
    \widetilde{\Sigma}_n = (\mathbb{I} \ot S^n) \widetilde{\Sigma}_0 \ , \ \ n=1,\ldots,d-1\
    .
\end{equation}
Again, the collection $\{ \widetilde{\Sigma}_0, \ldots ,
\widetilde{\Sigma}_{d-1}\}$ defines direct sum decomposition of
$\mathbb{C}^d \ot \mathbb{C}^d$
\begin{equation}\label{}
\mathbb{C}^d \ot \mathbb{C}^d  = \widetilde{\Sigma}_0 \oplus \ldots
\oplus \widetilde{\Sigma}_{d-1}\ .
\end{equation}
Moreover, operators $\widetilde{A}_n$ are defined as follows
\begin{eqnarray}\label{}
\widetilde{A}_n &=& \sum_{i,j=0}^{d-1}\, \widetilde{a}^{(n)}_{ij}\,
|i\>\<j| \ot S^n\, |\pi(i)\>\<\pi(j)|\, S^{n\, \dagger} \nonumber \\
&=& \sum_{i,j=0}^{d-1}\, \widetilde{a}^{(n)}_{ij}\, |i\>\<j| \ot
|\pi(i)+n\>\<\pi(j)+n|\ ,
\end{eqnarray}
with
\begin{equation}\label{a-tilde}
\widetilde{a}^{(n)} \, =\, \sum_{m=0}^{d-1}\, a^{(n+m)} \circ (\Pi
{S}^m)\ , \ \ \ \ \ \ \ \ (\mbox{mod $d$})\ ,
\end{equation}
where $\Pi$ is a permutation matrix corresponding to $\pi$, that is
\begin{equation}\label{}
    \Pi_{kl} = \delta_{k,\pi(l)}\ ,
\end{equation}
and $a \circ b$ denotes the Hadamard product of  $d\times d$
matrices $a$ and $b$, that is, $(a \circ b)_{ij} = a_{ij} b_{ij}$
\cite{Horn}.

\section{Entanglement witnesses for $d\otimes d$ systems}
\label{EWS}

Following \cite{iran} one constructs the following circulant
operators in $\mathbb{C}^d \ot \mathbb{C}^d$
\begin{equation}\label{}
    O_0 = \frac 1d\, \sum_{i=0}^{d-1} |ii\>\<ii|\
\end{equation}
and
\begin{equation}\label{}
    O_n = (\mathbb{I} \ot S^n) \, O_0 \, (\mathbb{I} \ot S^n)^\dagger =
     \frac 1d\, \sum_{i=0}^{d-1} |i,i+n\>\<i,i+n|\ ,
\end{equation}
for $n=1,\ldots,d-1$. Note that $O_\alpha$ defines a normalized,
i.e. ${\rm Tr} O_\alpha =1$, projector onto $\Sigma_\alpha$.
Finally, let $P^+_d$ denotes the projector onto the maximally
entangled states in $\mathbb{C}^d \ot \mathbb{C}^d$, that is
\begin{equation}\label{}
    P^+_d = \frac 1d \, \sum_{i,j=0}^{d-1} |ii\>\<jj|\ .
\end{equation}
Note that $O_m O_n = \frac 1d \delta_{mn} O_m$, and $O_m P^+_d = 0 $
for $n\geq 1$. Moreover,
\begin{equation}\label{OOO}
    O_0 + O_1 + \ldots + O_{d-1} = \frac 1d\, \mathbb{I}_d \ot
    \mathbb{I}_d\ .
\end{equation}
Let us consider the following family (parametrized by $\alpha$) of
circulant Hermitian operators
\begin{equation}\label{}
    W_\alpha = \mathbb{I}_d \ot
    \mathbb{I}_d - \frac 1\alpha O_1 - d(O_2 + \ldots O_{d-1}) -
    \left( 2 - \frac{1}{d\alpha} \right) P^+_d\ ,
\end{equation}
together with
\begin{equation}\label{}
    W'_\alpha = \mathcal{P}W_\alpha \mathcal{P}\ ,
\end{equation}
where $\mathcal{P}$ denotes permutation (flip/swap) operator defined
by
\begin{equation}\label{}
    \mathcal{P} = \sum_{i,j=0}^{d-1} |ij\>\<ji|\ .
\end{equation}
One easily finds
\begin{equation}\label{}
\mathcal{P} O_k \mathcal{P} = O_{\pi(k)}\ ,
\end{equation}
with $\pi$ defined in (\ref{pi}). Moreover, $\mathcal{P} P^+_d
\mathcal{P} = P^+_d$. Hence
\begin{equation}\label{}
    W'_\alpha = \mathbb{I}_d \ot
    \mathbb{I}_d - \frac 1\alpha O_{d-1} - d(O_1+ O_2 + \ldots O_{d-2}) -
    \left( 2 - \frac{1}{d\alpha} \right) P^+_d\ .
\end{equation}
 Note that for $d=3$ and $d=4$ one recovers Eqs. (17) and
(38), respectively, from Jafarizadeh et al. \cite{iran}. Note that
using (\ref{OOO}) one obtains a simplified formulae
\begin{equation}\label{}
    W_\alpha = dO_0 + \left(d - \frac 1\alpha \right)O_1 -
    \left( 2 - \frac{1}{d\alpha} \right) P^+_d\ ,
\end{equation}
and
\begin{equation}\label{}
    W'_\alpha = dO_0 + \left(d - \frac 1\alpha \right)O_{d-1} -
    \left( 2 - \frac{1}{d\alpha} \right) P^+_d\ .
\end{equation}
 Hence, up to a factor $\mu = 2 - \frac{1}{d\alpha}$,
$W_\alpha$ and $W'_\alpha$ belong to a class
\begin{equation}\label{aaa}
    \mathbf{W}[a_0,a_1,\ldots,a_{d-1}] = (a_0+1)O_0 + \sum_{n=1}^{d-1} a_i
    O_i - P^+_d\ ,
\end{equation}
that is,
\begin{equation}\label{our-W}
    W_\alpha = \mu \mathbf{W}[a_0,a_1,\ldots,a_{d-1}]\ ,
\end{equation}
with
\begin{eqnarray}
  a_0 &=& \frac d\mu - 1\, \nonumber \\
  a_1 &=& \frac d\mu - \frac{1}{\alpha\mu}\ , \\
  a_2 &=& a_3 = \ldots = a_{d-1} = 0 \nonumber \ .
\end{eqnarray}
Clearly, one obtains $W'_\alpha$ by interchanging $a_1$ and
$a_{d-1}$. Unfortunately, we do not know necessary and sufficient
conditions for $\mathbf{W}[a_0,a_1,\ldots,a_{d-1}]$ to be
entanglement witness. Clearly, $a_0,\ldots,a_{d-1}\geq 0$. Moreover,
one easily shows that necessarily
\begin{equation}\label{d-1}
    a_0 + a_1 + \ldots + a_{d-1} \geq d-1\ .
\end{equation}
Indeed, taking $\psi= \sum_{i=0}^{d-1} |i\> \in \mathbb{C}^d$, one
recovers (\ref{d-1}) from $\<\psi\ot
\psi|\mathbf{W}[a_0,a_1,\ldots,a_{d-1}]|\psi\ot\psi\> \geq 0$.

Finally, $\mathbf{W}[a_0,a_1,\ldots,a_{d-1}]$ becomes a positive
operator if and only if $a_0 \geq d-1$. Hence, necessarily
\begin{equation}\label{CP}
    a_0 < d-1\ .
\end{equation}
The problem is completely solved only for $d=3$ \cite{Kye}. In this
case apart from (\ref{d-1}) and (\ref{CP}) one has an additional
condition which says that if $a_0\leq 1$, then
\begin{equation}\label{}
    a_1 a_2 \geq (1-a_0)^2\ .
\end{equation}
Moreover, for $d=3$ we know that an entanglement witness
$\mathbf{W}[a_0,a_1,a_2]$ is non-decomposable if and only if
\cite{Kye}
\begin{equation}\label{nd}
a_1 a_2 < \frac{(2-a_0)^2}{4}\ .
\end{equation}
For $d>3$ we know only special cases when
$\mathbf{W}[a_0,a_1,\ldots,a_{d-1}]$ defines an EW. In particular it
is well known \cite{Osaka} (see also \cite{OSID,How}) that
$\mathbf{W}[d-2,1,0,\ldots,0]$ defines nd-EW.

Consider now a family of Hermitian operators defined in
(\ref{our-W}). Note that
\begin{equation}\label{}
    a_0 + a_1 + \ldots + a_{d-1} = a_0+a_1 = d-1\ ,
\end{equation}
and hence the necessary condition (\ref{d-1}) to be an EW is
satisfied. Now, to satisfy (\ref{CP}) one finds $\alpha > \frac
1d\,$. Let us observe that a convex combination
\begin{eqnarray}\label{}
 &&    a_1 \mathbf{W}[d-2,1,0,\ldots,0] + (1-a_1)
    \mathbf{W}[d-1,0,0,\ldots,0]  \nonumber \\ && =
    \mathbf{W}[(d-1)-a_1,a_1,0,\ldots,0]\ ,
\end{eqnarray}
defines an EW for any $a_1 \in (0,1]$. Now, the condition $a_1 \leq
1$ implies $\alpha \leq \frac{d-1}{d(d-2)}$. Hence, we have shown
that for any
\begin{equation}\label{MAIN}
\frac 1d \ < \ \alpha \ \leq \ \frac{d-1}{d(d-2)} \ ,
\end{equation}
the corresponding operator $W_\alpha$ defines an EW.

Let us observe that for $d=3$ formula (\ref{MAIN}) reproduces
analytical result $\alpha \in (\frac 13,\frac 23]$ from Jafarizadeh
et al. \cite{iran}. Note, however, that for $d=4$ the numerical
result $\alpha \in (\frac 14,\frac 13]$ from \cite{iran} is
improved. Formula (\ref{MAIN}) gives for $d=4$ the following
analytical condition $\alpha \in (\frac 14,\frac 38] \supset (\frac
14,\frac 13]$.

\section{A family of 2-qudit states}  \label{PPT}

In this section we show that whenever $\alpha$ satisfies
(\ref{MAIN}) then $W_\alpha$ and $W'_\alpha$ are non-decomposable
EW. In order to show it one has to construct a PPT entangled state
$\rho$ such that ${\rm Tr}(W_\alpha \rho)<0$. Consider the following
family of circulat 2-qudit states
\begin{equation}\label{rho}
    \rho = \sum_{i=1}^{d-1} \lambda_i O_i + \lambda_d P^+_d\ ,
\end{equation}
with $\lambda_n\geq 0$, and $\lambda_1 + \ldots +\lambda_{d-1} +
\lambda_d=1$. One easily finds
\begin{equation}\label{}
    {\rm Tr}(W_\alpha \rho) = (\lambda_1 - \lambda_d) \left( 1 -
    \frac{1}{d\alpha} \right)\ ,
\end{equation}
and
\begin{equation}\label{}
    {\rm Tr}(W'_\alpha \rho) = (\lambda_{d-1} - \lambda_d) \left( 1 -
    \frac{1}{d\alpha} \right)\ ,
\end{equation}
 Let us take the following special case corresponding to
\begin{eqnarray} \label{lll}
  \lambda_1 &=& \frac{ \beta}{\ell} \ , \nonumber \\
  \lambda_{d-1} &=&  \frac{(d-1)^2+1 -\beta}{\ell} \ ,  \\
  \lambda_d &=&  \frac{d-1}{\ell} \ . \nonumber
\end{eqnarray}
and $\lambda_2 = \ldots = \lambda_{d-2}  = \lambda_d$, with
\begin{equation}\label{}
    \ell = (d-1)(2d-3) +1\ .
\end{equation}
The parameter $\beta \in [0,(d-1)^2+1]$.  One has
\begin{equation}\label{}
    {\rm Tr}(W_\alpha \rho) =  \frac{ \beta - (d-1)}{\ell} \, \left( 1 -
    \frac{1}{d\alpha} \right)\ ,
\end{equation}
and
\begin{equation}\label{}
    {\rm Tr}(W'_\alpha \rho) =  \frac{ (d-1)(d-2) + 1 -  \beta}{\ell} \, \left( 1 -
    \frac{1}{d\alpha} \right)\ ,
\end{equation}
 It is easy to see that a state $\rho$ defined by (\ref{lll}) is
PPT iff $\lambda_1 \lambda_{d-1} \geq \lambda_d^2$ \cite{PPT-nasza}.
It gives the following condition for the parameter $\beta$
\begin{equation}\label{beta}
    1 \leq \beta \leq (d-1)^2\ .
\end{equation}
Note, that for $d=3$ formula (\ref{rho}) with $\lambda$s defined in
(\ref{lll})
\begin{equation}\label{}
    \rho = \frac 27 P^+_3 + \frac{\beta}{7} O_1 + \frac{5-\beta}{7}
    O_2\ ,
\end{equation}
reproduces  well know family of Horodecki states \cite{HOR}, and
(\ref{beta}) reproduces well known PPT condition: $1 \leq \beta \leq
4$. Actually, in this case a state is separable for $\beta \in
[2,3]$. It is PPT entangled for $\beta \in [1,2) \cup (3,4]$, and
NPT for $\beta \in [0,1) \cup (4,5]$. For $d=4$ our family
reproduces a state considered in \cite{iran} (see Eq. (43) with
$\gamma=3$).

It is clear that for $\beta < d-1$ and $\alpha$ satisfying
(\ref{MAIN}) $W_\alpha$ detects entanglement  of $\rho$ which proves
that $W_\alpha$ is non-decomposable EW. Similarly, for $\beta >
(d-1)(d-2)+1$ and $\alpha$ satisfying (\ref{MAIN}) $W'_\alpha$
detects entanglement of $\rho$ which proves that $W'_\alpha$ is
non-decomposable EW. As a byproduct we showed that for $$\beta \in
[1,d-1) \ \cup\  ((d-1)(d-2)+1,(d-1)^2+1]$$ a state $\rho$ is PPT
entangled.

\section{Conclusions}

We provide a new class of non-decomposable entanglement witnesses
for $d \ot d$ systems (two qudits). Our construction generalizes the
one proposed recently by Jafarizadeh et al. \cite{iran} for $d=3$
and $d=4$ on the basis of semidefinite linear programming. We stress
that for $d=3$ we recover analytical result of \cite{iran}. However,
for $d=4$ our analytical result slightly improves numerical result
of \cite{iran}. As a byproduct, we provided a new class of PPT
entangled states detected by our witnesses which generalizes well
known family of states constructed by Horodecki et al. for $d=3$
\cite{HOR}.

For the experimental realization of  entanglement witnesses
$W_\alpha$ and $W_\alpha'$ discussed in this paper one can use for
example the generalized Gell-Mann matrix basis (see e.g.
\cite{Kimura,BK}) consisting in the following set of Hermitian
matrices: symmetric
\begin{equation}\label{GI}
    \Lambda_s^{k\ell} = |k\>\<\ell| + |\ell\>\<k| \ , \ \ 0\leq k < \ell \leq
    d-1 \ ,
\end{equation}
antisymmetric
\begin{equation}\label{GII}
    \Lambda_a^{k\ell} = -i|k\>\<\ell| + i|\ell\>\<k|) \ , \ \ 0\leq k < \ell \leq
    d-1 \ ,
\end{equation}
and diagonal
\begin{equation}\label{GIII}
    \Lambda^{\ell} = \sqrt{\frac{2}{\ell(\ell+1)}} \left( \sum_{j=0}^{\ell-1}|j\>\<j| - (\ell-1)|\ell\>\<\ell|\right)  \
    ,
\end{equation}
for $1\leq \ell \leq d-2$. For $d=3$ the above set reconstructs the
standard $3\times 3$ Gell-Mann matrices. Let us observe that in this
case Gell-Mann matrices may be defined in terms of spin-1 operators
\begin{equation}\label{}
S_{x}=\frac{\hbar}{\sqrt{2}}\left(\begin{array}{ccc}
0 & 1 & 0\\
1 & 0 & 1\\
0 & 1 & 0\end{array}\right),\quad
S_{y}=\frac{\hbar}{\sqrt{2}}\left(\begin{array}{ccc}
0 & -i & 0\\
i & 0 & -i\\
0 & i & 0\end{array}\right),
\end{equation}
and
\begin{equation}\label{}
 S_{z}=\frac{\hbar}{\sqrt{2}}\left(\begin{array}{ccc}
1 & 0 & 0\\
0 & 0 & 0\\
0 & 0 & -1\end{array}\right)\ .
\end{equation}
One finds the following relations \cite{BK}
\begin{eqnarray*}
 \Lambda^{01}_s  &=& \frac{1}{\sqrt{2}\hbar^{2}}\left(\hbar
S_{x}+\left\{ S_{z},S_{x}\right\}
\right) \\
 \Lambda^{02}_s  &=& \frac{1}{\hbar^{2}}\left(S_{x}^{2}-S_{y}^{2}\right), \\
  \Lambda^{12}_s &=& \frac{1}{\sqrt{2}\hbar^{2}}\left(\hbar
S_{x}-\left\{ S_{z},S_{x}\right\} \right), \\
  \Lambda^{01}_a &=& \frac{1}{\sqrt{2}\hbar^{2}}\left(\hbar
S_{y}+\left\{ S_{y},S_{z}\right\}
\right), \\
  \Lambda^{02}_a &=& \frac{1}{\hbar^{2}}\left\{
S_{x},S_{y}\right\} \\
  \Lambda^{12}_a &=& \frac{1}{\sqrt{2}\hbar^{2}}\left(\hbar
S_{y}-\left\{ S_{y},S_{z}\right\} \right), \\
  \Lambda^{0} &=& 2\mathbb{I}_3+\frac{1}{\sqrt{2}\hbar^{2}}\left(\hbar
S_{z}-3S_{x}^{2}-3S_{y}^{2}\right), \\
  \Lambda^{1} &=& \frac{1}{\sqrt{3}}\left(-2\mathbb{I}_3+\frac{3}{2\hbar^{2}}\left(\hbar
S_{z}+S_{x}^{2}+S_{y}^{2}\right)\right),
\end{eqnarray*}
where $\{A,B\}= AB+BA$. Finally, one finds the following
representation of  $W_\alpha$ in terms of local Hermitian operators
\begin{eqnarray}
W_{\alpha} & = & \left(4-\frac{2}{3\alpha}\right) \mathbb{I}_3 \ot
\mathbb{I}_3 - \left(\frac{1}{3}-\frac{5}{9\alpha}\right)
\left[\Lambda^{0}\otimes\Lambda^{0} \right. \nonumber
\\ &+& \left.
\Lambda^{1}\otimes\Lambda^{1} \right]+
\sqrt{3}\left(1-\frac{1}{3\alpha}\right) \left[
\Lambda^{0}\otimes\Lambda^{1} - \Lambda^{1}\otimes\Lambda^{0}\right]
 \nonumber \\
&-&\left(\frac{4}{3}-\frac{2}{9\alpha}\right)
  \left[
  \Lambda_{s}^{01}\otimes\Lambda_{s}^{01}   +   \Lambda_{s}^{02}\otimes\Lambda_{s}^{02} +
  \Lambda_{s}^{12}\otimes\Lambda_{s}^{12} \right.
  \nonumber \\ &-& \left.
  \Lambda_{a}^{01}\otimes\Lambda_{a}^{01}  -
  \Lambda_{a}^{02}\otimes\Lambda_{a}^{02} -
  \Lambda_{a}^{12}\otimes\Lambda_{a}^{12} \right].
  \end{eqnarray}
Similar decomposition can be found for $W_\alpha'$. Hence,  to
measure entanglement witnesses $W_\alpha$  and $W_\alpha'$ one can
perform suitable number of purely local measurement settings
$\Lambda^\mu \ot \Lambda^\nu$. Using the general scheme
(\ref{GI})-(\ref{GIII}) one can represent $W_\alpha$ and $W_\alpha'$
for arbitrary (finite) $d$ as a combination of purely local
observables. In this way one can experimentally find out whether a
given 2-qudit state is entangled or not.

\section*{Acknowledgments}

This work was partially supported by the Polish Ministry of Science
and Higher Education Grant No \\ 3004/B/H03/2007/33.

\end{document}